# Physics Archives

*November 2009*

*Adaptive BLASTing through the Sequence Dataspace:*
*Theories on Protein Sequence Embedding*

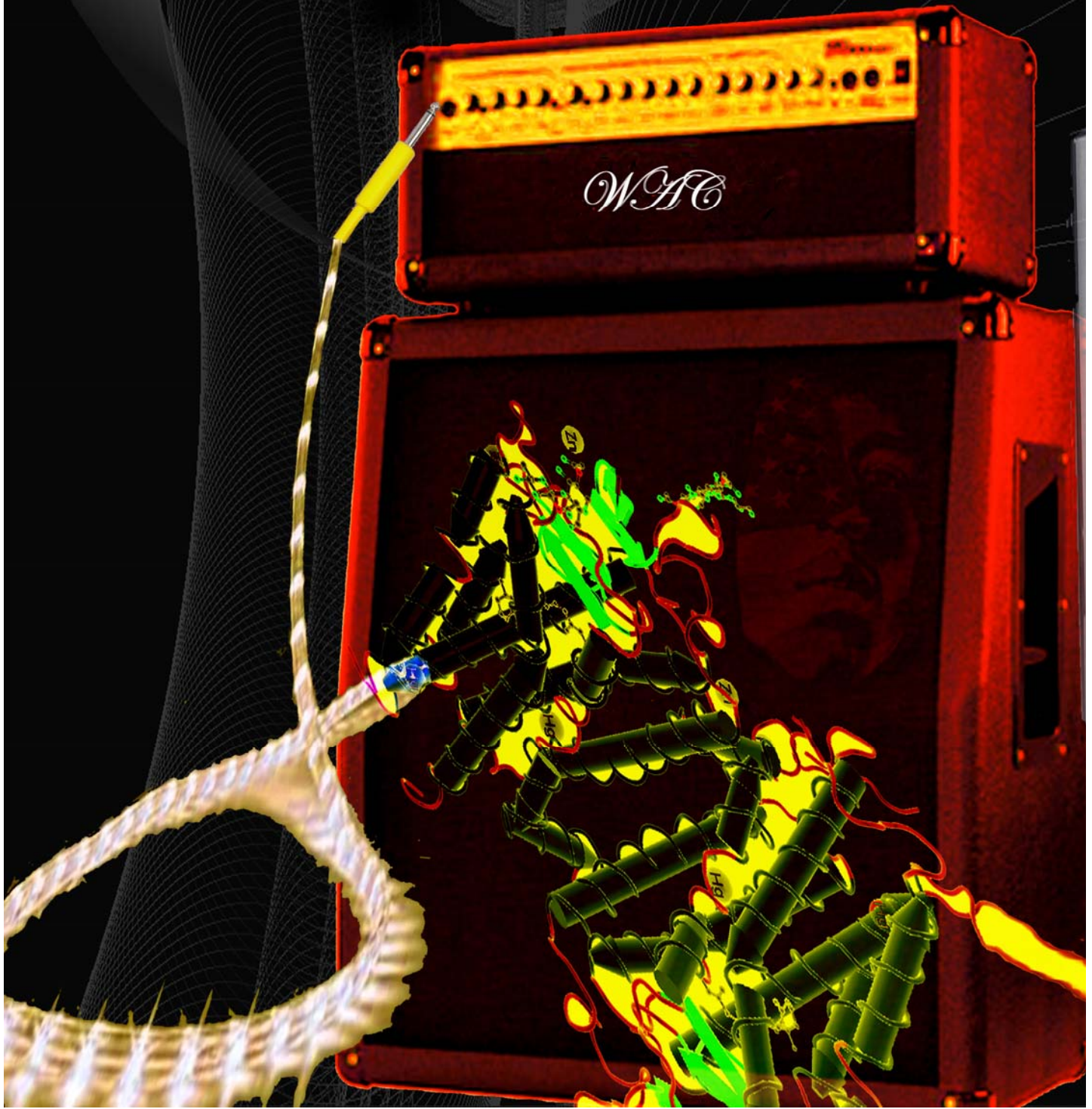

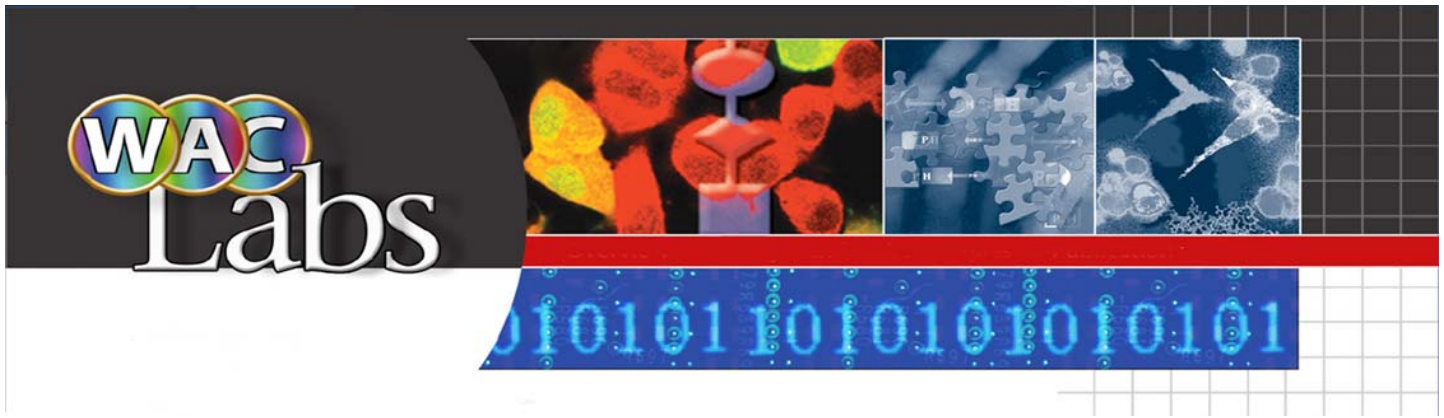

# Adaptive BLASTing through the Sequence Dataspace: Theories on Protein Sequence Embedding

Yoojin Hong [1,2], Jaewoo Kang [3,4*], Dongwon Lee [1,5], Randen L. Patterson [2,6], and Damian B. van Rossum [2,6*]

(1) Department of Computer Science and Engineering, The Pennsylvania State University, USA
(2) Center for Computational Proteomics, The Pennsylvania State University, USA
(3) Department of Computer Science and Engineering, Korea University, Korea
(4) Department of Biostatistics, College of Medicine, Korea University, Korea
(5) College of Information Sciences and Technology, The Pennsylvania State University, USA
(6) Department of Biology, The Pennsylvania State University, USA

*To whom correspondence should be addressed:
kangj@korea.ac.kr; dbv10@psu.edu

A major computational challenge in the genomic era is annotating structure/function to the vast quantities of sequence information now available.  This problem is illustrated by the fact that most proteins lack comprehensive annotation, even when experimental evidence exists.  We theorized that phylogenetic profiles provide a quantitative method that can relate the structural and functional properties of proteins, as well as their evolutionary relationships.  A key feature of phylogenetic profiles is the interoperable data format (e.g. alignment information, physiochemical information, genomic information, etc).  Indeed, we have previously demonstrated Position Specific Scoring Matrices (PSSMs) are an informative M-dimension which can be scored from quantitative measure of embedded or unmodified sequence alignments.  Moreover, the information obtained from these alignments is informative, even in the "twilight zone" of sequence similarity (<25% identity)(1-5).  Although powerful, our previous embedding strategy suffered from contaminating alignments (embedded AND unmodified) and computational expense.  Herein, we describe the logic and algorithmic process for a heuristic embedding strategy (Adaptive GDDA-BLAST, Ada-BLAST). Ada-BLAST on average up to ~19-fold faster and has similar sensitivity to our previous method.  Further, we provide data demonstrating the benefits of embedded alignment measurements for isolating secondary structural elements and the classifying transmembrane-domain structure/function. Together, these advances allow for further exploration of the embedded alignment data space within sufficiently large data sets such that relevant statistical inferences can be achieved.  We theorize that sequence-embedding is one of multiple ways that low-identity alignments can be measured and incorporated into high-performance PSSM-based phylogenetic profiles.

**Introduction**

One of the major challenges that biologists face is the ability to identify relationships between highly divergent protein sequences. Although many methods (e.g., (2;6;7)) have addressed the problem, it still remains as a challenge, as conventional sequence alignment methods often fail to obtain statistically robust measurements when sequence identity dips into the "twilight zone" (~ less than 25% identity). In general, when pairwise sequence alignments between protein sequences fall below 25% identity, statistical measurements do not provide support robust enough to identify clear phylogenetic relationships, structural features, or protein function despite intensive research in this area (2;8-11). GDDA-BLAST (Gestalt Domain Detection Algorithm - Basic Local Alignment Search Tool), originally introduced in (12), was designed to address the challenges associated with low-identity alignments/divergence. We determined that this alignment information, when incorporated into phylogenetic profiles, is informative to our laboratory experiments at multiple scales (e.g. whole protein, single protein domain, and single amino acid)(3-5;13-16). We have used these analyses: (i) to reconstruct evolutionary histories, (ii) to identify functions in domains of unknown function, (iii) to classify structural homologues of high sequence divergence, and (iv) to inform our biochemical experimentation by isolating key amino acids important to protein function.

A phylogenetic profile of a protein is a vector, where each entry quantifies the existence of the protein in a different genome. This approach has been shown to be applicable to whole molecule (Single Profile Method), to an isolated domain (Multiple Profile Method), and to individual amino acids(17-19). GDDA-BLAST matrices are a variation of phylogenetic profiles, except in our case, a protein is a vector where each entry quantifies the existence of alignments with a PSSM(1;2). The basic idea underlying our method begins by compiling a set of PSSMs that the query sequence is compared to. These profiles can be obtained from any protein-sequence knowledge-base source (e.g. Protein Data Bank, Pfam, SMART, NCBI Conserved Domain Database (CDD)(20-23)), or generated locally using PSI-BLAST(24). We employ reverse specific position BLAST (rps-BLAST (23)) to compare query and PSSMs, and have introduced multiple innovations in GDDA-BLAST. We utilize a single domain PSSM database for pairwise comparisons. Then we record and quantify all alignments between an unmodified (control), and modified query sequence. The latter is composed of two types of alignments: "seeded" and non-seeded alignments. We modify the query sequence with a "seed" from the PSSM, creating a consistent initiation site (Figure 1a-b). "Seeds" are generated from profiles by taking a portion (e.g. 10% in this study) of the PSSM sequence (e.g. from the N-terminus or C-terminus). These thresholds were used based on results from our previous studies. These "seeds" are embedded at each position of the query sequence. For example, a query sequence of 100 amino acids yields 100 distinct test sequences for each "seed". This strategy was designed to amplify and encode the alignments possible for any given query sequence. Instead of a sliding window, we utilized a sliding "seed", similar yet inverse to the embedding strategies employed by Henikoff and Henikoff(25).

Each of these modified query sequences is then aligned by rps-BLAST against the parent profile. Since BLAST algorithms are based on a "hit and extension of the hit" approach, the embedded "seed" creates consistent initiation site which allows rps-BLAST to extend an alignment even between highly divergent sequences (Figure 1c). Next, we filter our results from rps-BLAST using thresholds of # of hits, percentage identity and percentage coverage (i.e. alignment length as a function of profile length) (Figure 1d). The output of these comparisons is a composite [product] score of either zero [when there is no significant match] or a positive value [which measures the degree of successful matching of the protein sequence to each of the profiles]. This procedure can be readily adapted to make an unbiased comparison between a series of query sequences by subjecting them to the same screening analysis with the same set of PSSM sequences as "seeds". From these results, each profile alignment above threshold is defined within the query sequence to create boundaries for our subsequent pairwise alignments. Residue scores for each amino acid position are calculated, in the absence of the embedded sequence, by scoring a value=2 for identities and

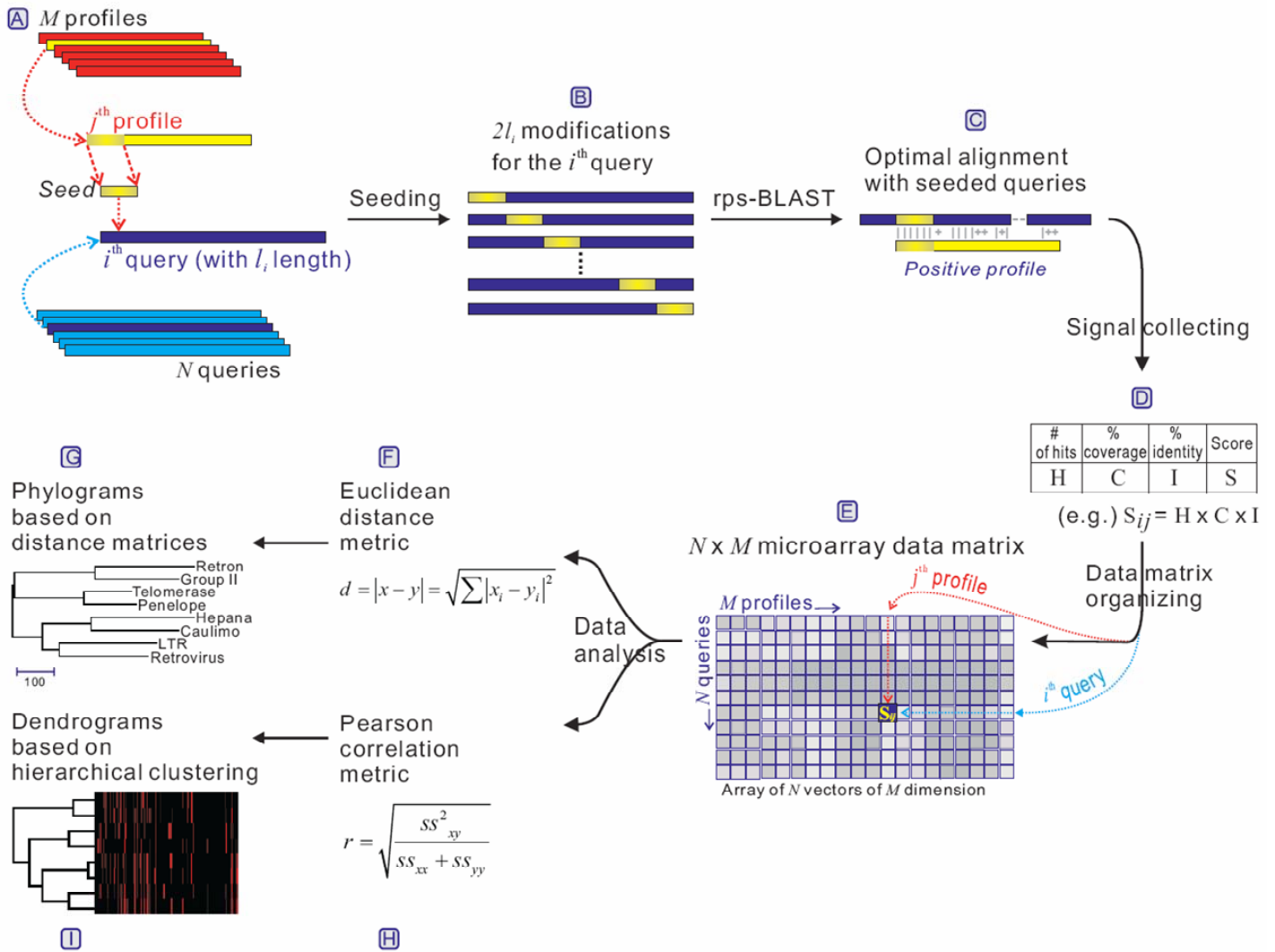

**Figure 1 GDDA-BLAST Concept.** This schematic depicts the work flow of GDDA-BLAST (A-B) The algorithm begins with a modification of the query amino acid sequence at each amino acid position via the insertion of a "seed" sequence from the profile of interest. These seeds are obtained from the profile consensus sequences from NCBI's Conserved Domain Database (CDD). (C-E) Signals are collected from optimal alignments between the "seeded" sequences and profiles using rps-BLAST and are incorporated as a composite score into an N by M data matrix. (F-I) This data space can be analyzed to generate phylograms and dendrograms based on Euclidean distance measures and Pearson correlation measures of GDDA-BLAST signals, respectively.

value=1 for positive substitutions for all alignments above threshold. These scores are then summed from each alignment, providing a total residue score. Once scored, each query sequence (N) is represented in a vector of non-negative numbers in M dimensions (M= # of "seeds" tested) (Figure 1e). This NxM data matrix can then be used to create a tree of relationships based on standard statistical techniques such as hierarchical clustering based on Euclidean distance or Pearson's correlation between each query sequence (Figure 1f-i).

Despite the great potential of embedded alignment strategies for answering a diverse set of biological questions, its computational cost is prohibitively expensive. This is due to the current method for generating and analyzing embedded sequences. As proteins range in length from tens of amino acids to < 8000, proteomic scale studies using GDDA-BLAST are prohibitive. To address this challenge we present here a novel sequence alignment algorithm that is as sensitive as GDDA-BLAST but orders of magnitude faster. Ada-BLAST (Adaptive GDDA-BLAST) is termed for its adaptive nature, and exploits the similarity among embedded sequences to adaptively avoid expensive computations.

## Methods
### Definitions

Let the target sequence be $X$ and the query sequence be $Y$. Note that the sequence profile, to measure any query protein sequence, is called the target sequence. The length of a sequence $X$ is denoted as $|X|$. Assume that $|X|$ and $|Y|$ are $n$ and $m$, respectively. A subsequence of $X$ from the $i$-th residue to the $j$-th residue is denoted by $x_{i,j}$ such that $0 \le i \le j \le n-1$. A subsequence of length one such as $x_{i,i}$ is simply represented as $x_i$. Concatenation of two sequences, $X$ and $Y$, is represented as $X|Y$. Two subsequences that are aligned in an alignment are represented with $()$. For example, $(x_{a,b}, y_{c,d})$ represents that $x_{a,b}$ and $y_{c,d}$ are aligned.

|  |  |  |  | $y_{0,q-1}$ |  | N-terminal seed (S) |  |  | $y_{q,m-1}$ |  |  |
|---|---|---|---|---|---|---|---|---|---|---|---|
| Residue | $y_0$ | $y_1$ | ... | $y_{q-1}$ | $x_0$ | $x_1$ | ... | $x_{k-1}$ | $y_q$ | $y_{q+1}$ | ... | $y_{m-1}$ |
| Chimera index | $c_0$ | $c_1$ | ... | $c_{q-1}$ | $c_q$ | $c_{q+1}$ | ... | $c_{q+k-1}$ | $c_{q+k}$ | $c_{q+k+1}$ | ... | $c_{m+k-1}$ |

**TABLE 1. Residues of a chimera sequence.** This shows an example chimera sequence with an N-terminal seed of length $k$ inserted into the position $q$ of original query sequence Y. The length of the resulting chimera sequence is $m+k$ where $m$ is the length of the original query sequence.

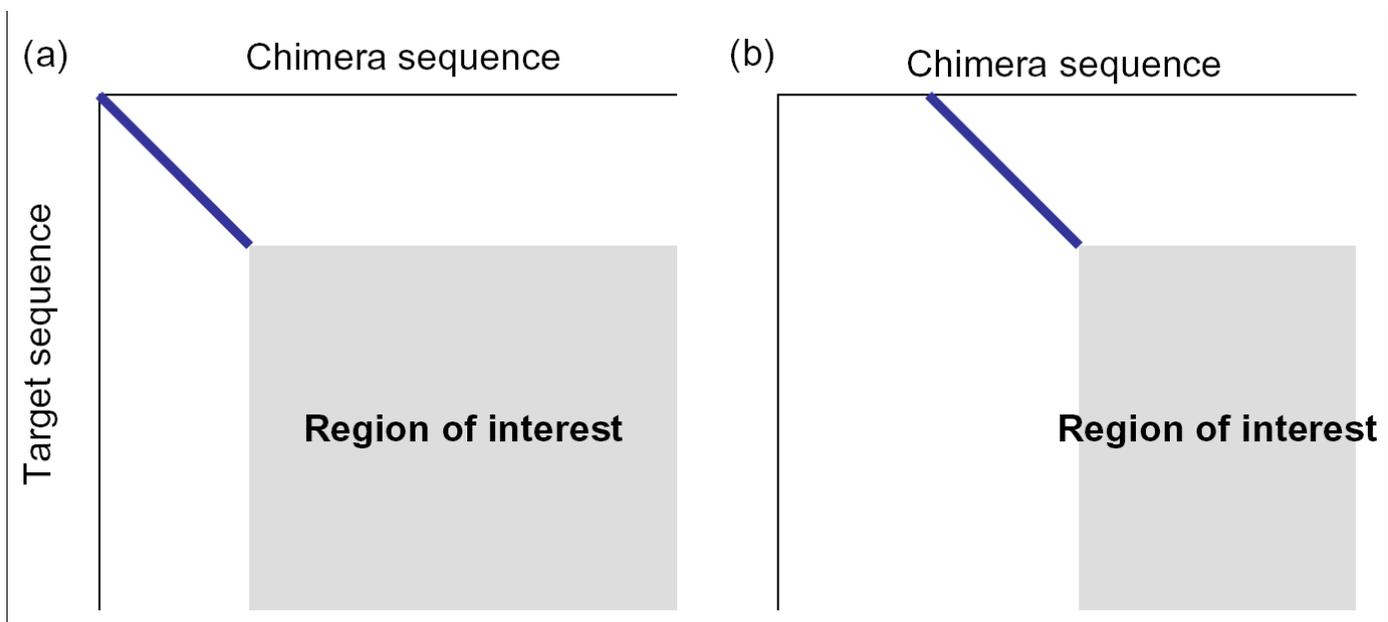

**Figure 2 Limiting the region of interest with respect to the seed insertion position.** The diagonal line represents the alignment with seed in different locations. The examples illustrate the region of interest for N-terminal seeds. Similarly for C-terminal seed, it is the upper-left corner of the seed.

An embedded (chimera) sequence is generated by embedding either the N- or C-terminal portion of $X$ as a "seed", denoted by $S$, to every position of $Y$ (Figure 2). Usually $p\%$ of $X$ (i.e., $k$ residues of $X$ where $k = \lceil |X| \times p \times 0.01 \rceil$) is used as a seed. Thus, N- and C-terminal seeds are $x_{0,k-1}$ and $x_{n-k,n-1}$, respectively. A chimera with a seed at position $q$ of $Y$ is $y_{0,q-1}|S|y_{q,m-1}$ and represented with $C(q)$. An example of a chimera is shown in Table 1. To align the target sequence $X$ and the query sequence $Y$, GDDA-BLAST generates $(m \times 2)$ number of chimera sequences inserting N- and

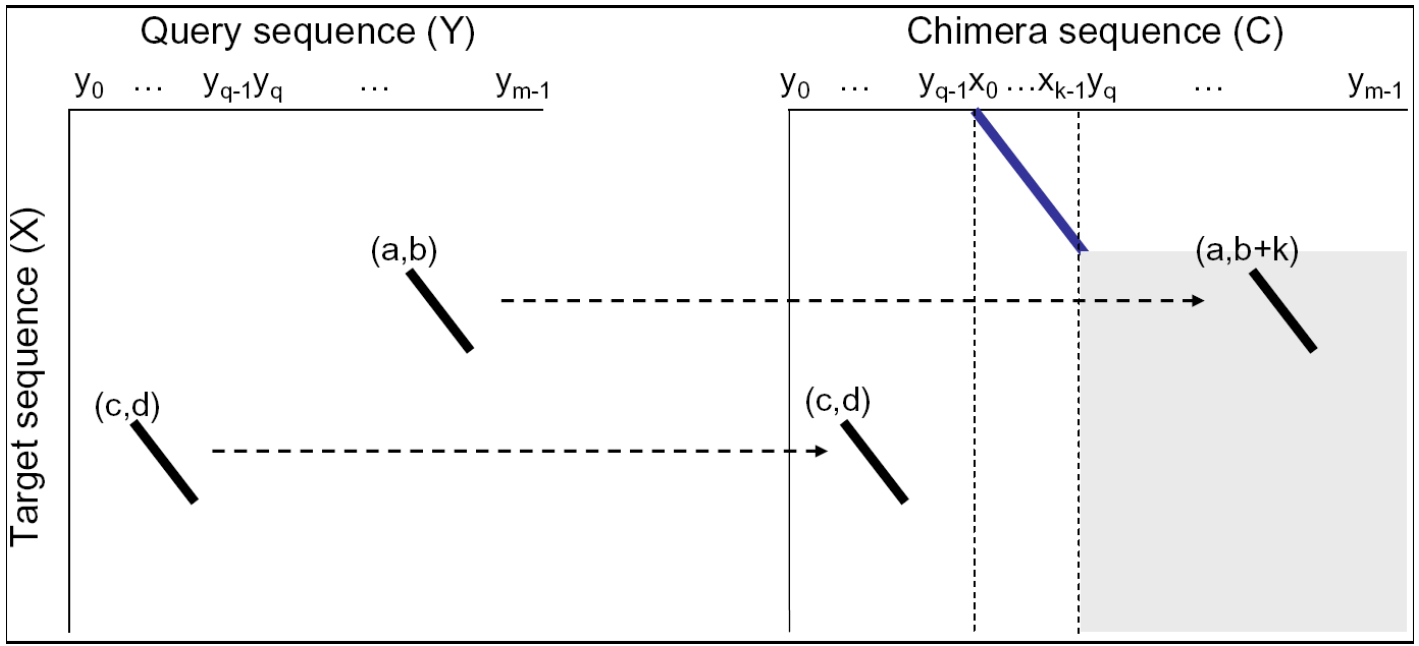

**Figure 3 Corresponding hits between a query and a chimera sequence.** This example illustrates the hits between the target sequence (X) and the query sequence (Y) can be reused for aligning a chimera sequence (C) against the target sequence (X).

C-terminal seeds from $X$ at each position of $Y$ (Figure 3). Each chimera is then aligned to $X$ using rps-BLAST. For each alignment, rps-BLAST is run independently, yielding a total of $(m \times 2 \times$ # of target sequences$)$ BLAST executions.

```
Seed Position : 292
Query: 293  LVDCSDNSANIKE-------PVPEPANAGKRKVREFNFEKWNARITDLRKQVEELFERKY 345
            LVDCSDNSANIKE        P+P P + G              LR+ ++ L+
Sbjct: 1    LVDCSDNSANIKEVLLSPCDPLPCPDHRGGNYNLSVTGT--------LREDIKSLYVDLA 52

Query: 346  AQAIKAKGPVTIPYPLFQSHVEDLYVEGLPEGIPFRRPSTYGIPRLERI 394
                + K       F              P    P R+  Y    +  +
Sbjct: 53   LMSQGIKVLNPDNSYDFCE-----AGLPKPSFCPLRKGQQYSYAKTVNV 96

Seed Position : 293
Query: 294  LVDCSDNSANIKEV--------PEPANAGKRKVREFNFEKWNARITDLRKQVEELFERKY 345
            LVDCSDNSANIKEV           P P + G              LR+ ++ L+
Sbjct: 1    LVDCSDNSANIKEVLLSPCDPLPCPDHRGGNYNLSVTGT--------LREDIKSLYVDLA 52

Query: 346  AQAIKAKGPVTIPYPLFQSHVEDLYVEGLPEGIPFRRPSTYGIPRLERI 394
                + K       F              P    P R+  Y    +  +
Sbjct: 53   LMSQGIKVLNPDNSYDFCE-----AGLPKPSFCPLRKGQQYSYAKTVNV 96

Seed Position: 294
Query: 295  LVDCSDNSANIKE---------PEPANAGKRKVREFNFEKWNARITDLRKQVEELFERKY 345
            LVDCSDNSANIKE            P P + G              LR+ ++ L+
Sbjct: 1    LVDCSDNSANIKEVLLSPCDPLPCPDHRGGNYNLSVTGT--------LREDIKSLYVDLA 52

Query: 346  AQAIKAKGPVTIPYPLFQSHVEDLYVEGLPEGIPFRRPSTYGIPRLERI 394
                + K       F              P    P R+  Y    +  +
Sbjct: 53   LMSQGIKVLNPDNSYDFCE-----AGLPKPSFCPLRKGQQYSYAKTVNV 96
```

**Figure 4 rps-BLAST alignments for three consecutive chimera sequences.** The query and the target sequences are general transcription factor II, i isoform from *Homo sapiens* (NP001509.2) and ML(MD2related lipidrecognition) domain (cd00912), respectively.

Note that chimera sequences are different *only* by the position of the seed (Figure 4). This implies that for two subsequent chimeras, much of the computation can be reused. In the Ada-BLAST approach, we re-use outputs from each step of rps-BLAST for efficient computation.

For clarity, we define the outputs of each step as follows. In the first step of rps-BLAST, we find hits between $X$ and $Y$. A hit of $(x_{i,i+w-1}, y_{j,j+w-1})$, where $w$ is a word size, is denoted as $h(i, j)$ (Figure 5a). After ungapped extension on two neighboring hits in the second step, we obtain HSPs(High Scoring Sequence Pairs) extending two hits without gaps. An HSP to align $x_{i,i+r}$ and $y_{j,j+r}$, is denoted as $hsp(i,j,r)$ (Figure 5b). If an HSP has a score high enough to trigger gapped extension, in the third step, an alignment is generated extending the HSP with gaps to both directions from a residue pair in the highest scored region of the HSP (Figure 5c-d). Note that the pair from which gapped extension is started is also referred to as a seed (24). In order to avoid confusion, we denote this as a *GE starting pair* to distinguish if from the embedded seed of GDDA-BLAST.

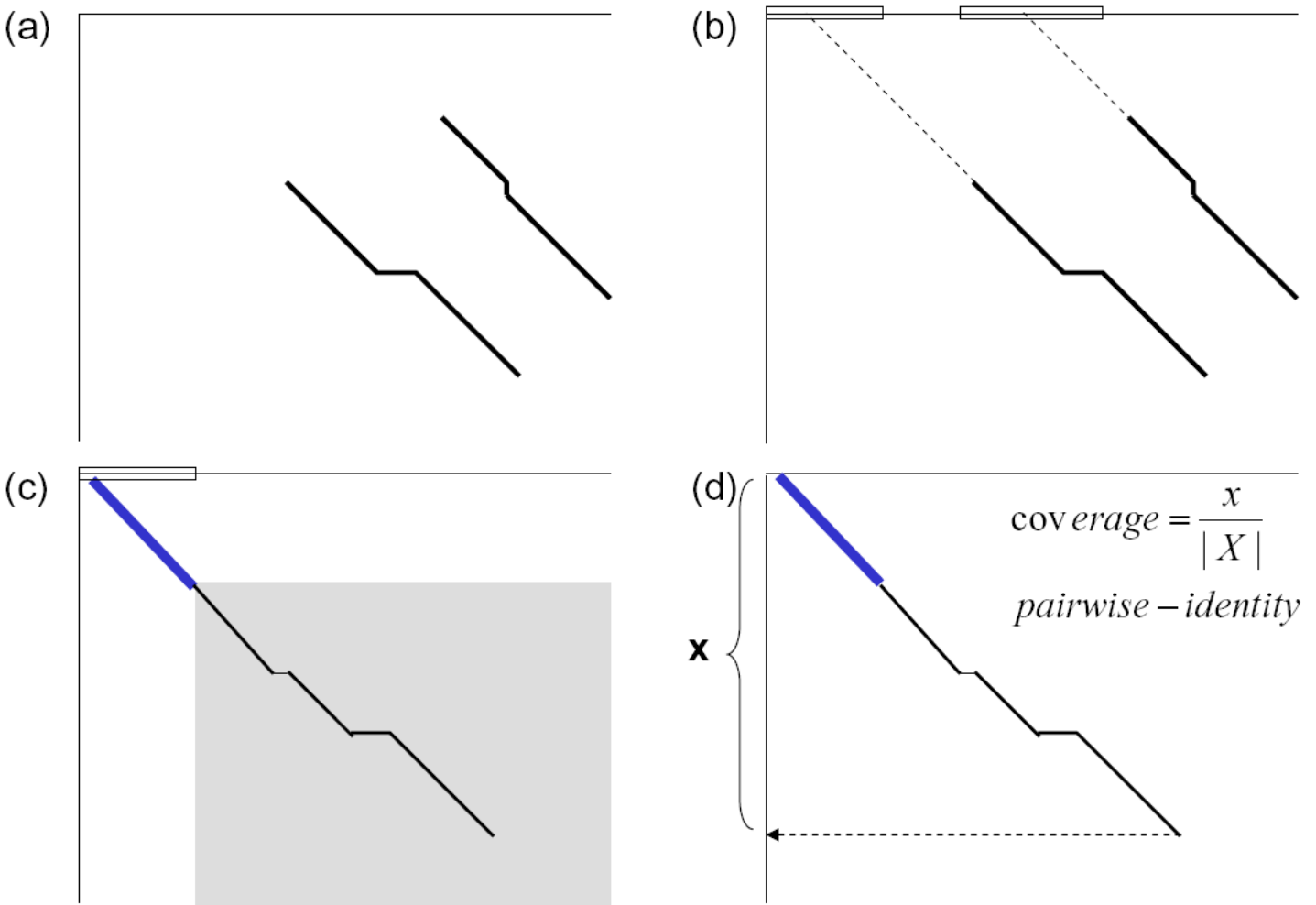

**Figure 5 Four basic steps of Ada-BLAST.** (a) Step 1: Find multiple non-overlapping local alignments (b) Step 2: Select seed insertion positions in query sequence (c) Step 3: Generate final alignments with seed (d) Step 4: Filter out non-significant alignments using coverage and pairwise identity of the alignment.

$$coverage = \frac{x}{|X|}$$

$$pairwise-identity$$

## Observations of GDDA-BLAST

**Observation 1.** *Seeding limits the search space.* Since a seed provides an exact match, it is very likely that the GE starting pair is in an HSP including the seed. Also, we are only interested in the alignments including the seed because the other alignments can be found through conventional methods using the original query sequence. This limits the search space of rps-BLAST. For example, when a seed is inserted at the position 0 of a query sequence, our search space will be the region in gray (Figure 2a), and every time the seed insertion position is moved to the right, our search space is reduced as shown in Figure 2b. Note that, in the case of the chimeras with C-terminal seed, the search space is limited to the upper-left corner from the start position of the seed.

**Observation 2.** *Chimeras share hits.* Because chimeras are the same sequence except for the position of a seed, most of their hits are conserved (Figure 6). Therefore, we can reuse hits between

X and Y to compute the alignments for any chimera sequences. Consider a chimera, C(q). Let $h_{C(q)}$ be a hit obtained after the first rps-BLAST step between X and C(q). The relation between X-Y hits (i.e., $h$) and X-C(q) hits (i.e., $h_{C(q)}$) can be defined as follows:

*Lemma 1.*

$$h(a,b) = \begin{cases} h_{C(q)}(a,b) & \text{if } 0 \leq b \leq q-w \\ h_{C(q)}(a,b+k) & \text{if } q \leq b \leq m \\ no\,corresponding\,hit & otherwise \end{cases}$$

where $k$ is the seed length, $q$ is the seed insertion position, and $w$ is the size of a hit.
Proof is omitted because it is obvious from Figure 4. Note that we are *not* interested in the hits in case Figure 2a because they are out of the region of our interest by observation 1. Therefore, we only use hits in case Figure 2b for alignment.

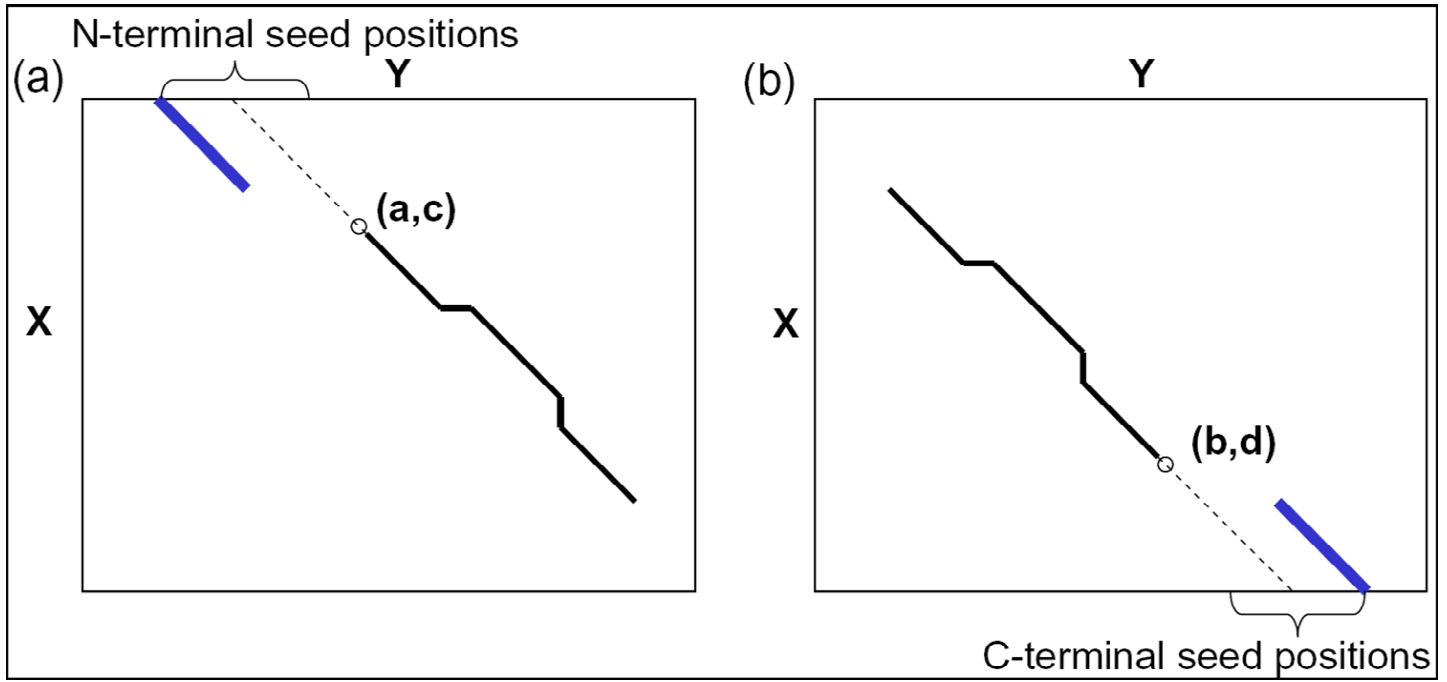

**Figure 6 Selecting seed positions given a partial alignment.** Ranges on the top and bottom represent seed insertion positions, and X and Y are a target and a query sequence, respectively. (a) N-terminal seed positions (b) C-terminal seed positions.

*Observation 3. Chimeras share HSPs after ungapped extension.* rps-BLAST performs ungapped extension on neighboring hits resulting in HSPs. Similar to observation 2, we can define the relationship between an HSP of X-Y (i.e., $hsp$) and that of X-C(q) (i.e., $hsp_{C(q)}$) as follows:

*Lemma 2.*

$$hsp(a,b,r) = \begin{cases} hsp_{C(q)}(a,b,r) & \text{if } 0 \leq b \leq q-r \\ hsp_{C(q)}(a,b+k,r) & \text{if } q \leq b \leq m \\ no\,corresponding\,HSP & otherwise \end{cases}$$

where $r$ is the length of the HSP. The proof is straightforward from lemma 1.

***Observation 4.*** *Chimeras share alignment paths in gapped extension.* Gapped extension in rps-BLAST starts at a *GE starting pair* that is a central residue pair in the highest scoring segment of any HSP whose score is sufficiently high. Different alignments can be generated if gapped extension is performed from different GE starting pairs and there is no guarantee that the same GE starting pair is selected for different chimera sequences. However, as shown in Figure 4 (boxed residues represent the seed), if a portion of a target sequence is conserved in a query sequence, it is very likely that the conserved region is aligned for multiple neighboring chimera sequences. We exploit this property to speed up the alignment process.

***Observation 5.*** *Not every chimera produces a useful alignment.* Even though a seed provides artificial matches, it cannot be extended if there are insufficient HSPs nearby to connect to. Therefore, we can significantly reduce the computational complexity of the alignment process by inserting seeds only to a limited number of query positions that are likely to be extended. Hence, in Ada-BLAST, we align the query and the target sequence first and then compute the seed insertion positions from the alignment result before aligning the chimera sequences.

### Ada-BLAST Algorithm

Ada-BLAST works in four basic steps as shown in Figure 5. First, we find conserved regions by generating non-overlapping local alignments between the query and the target sequence(26). We call these *partial alignments*. Second, for each partial alignment from step 1, seed insertion positions are determined. Third, we produce *final alignments* including the seeds. Finally, we filter out non-significant alignments using quality measures such as the % coverage and % identity of the alignment to the corresponding PSSM.

***Step 1. Find multiple non-overlapping local alignments.*** Haung et al (26) proposed an algorithm to find multiple non-overlapping local alignments between two sequences. We adopted this algorithm to generate partial alignments between the query and the target sequence. For any target sequence, we can use any given substitution matrix (e.g. BLOSUM62, BLOSUM45, PAM30, etc) in our algorithm.

The local alignments are found as follows. First, hits between the query and the target sequence are found. A hit is three consecutive residues with score larger than a threshold (i.e. *minimum word score*). For each hit, to generate an HSP, ungapped extension is performed until the score drops below a threshold (*HSP drop-off score*). When a new HSP is constructed, hits involved in the HSP are removed to prevent subsequent HSPs extending over them. This ensures that all local alignments produced later are non-overlapping. We keep only the HSPs with score larger than a threshold (*minimum HSP score*). Gapped extension is then performed for each HSP to generate partial alignments.

We keep only the partial alignments whose lengths are greater than threshold (i.e. *minimum partial alignment length*). *Minimum partial alignment length* is determined proportionally to the length of the target sequence. Since a partial alignment is a locally best alignment, it is not likely that a seed will be extended further than the end position of the partial alignment. In the final step, a final alignment will be filtered by the coverage of the alignment over a target sequence. This pre-filtering on partial alignments can remove seed insertion positions where a seed cannot be extended to final alignments with sufficient coverage.

***Step 2. Select seed insertion positions.*** In this step, seed insertion positions in the query sequence are selected given the partial alignments obtained from step 1. As discussed in observation 1, a final alignment is generated by extending a seed from its end positions. Since a partial alignment is a locally optimal alignment, the extension of a seed can be converged to the partial alignment if a seed is inserted nearby and the score of the path is high enough. Because of the relatively high penalty of gaps used in sequence alignment methods, an alignment is usually generated with HSPs connected with small numbers of gaps in between. In addition, the score of partial alignment on either side of the gaps must be higher than the gap penalty [17]. The gapped extension usually starts from the seed to the partial alignment because the score of the alignment with the seed is typically much higher than that of the partial alignment. For this reason, we can compute the seed insertion positions simply with the score of a seed and the distance from the seed to the partial alignment.

Given a seed $S$ of score $Score(S)$, the maximum gap $G(S)$ (i.e., the distance from the seed to the partial alignment) is computed as follows: $G(S) = \lceil \frac{Score(S) - GOP}{|GEP|} \rceil - 1$, where $GOP$ and $GEP$ are gap opening penalty and gap extension penalty, respectively. Given a query sequence Y and a partial alignment ($x_{a,b}$, $y_{c,d}$), the query position $q$ is subject to embedding a seed of length $k$ as follows: (1) for N-terminal seed: $\max[-k, \varepsilon - G(S)] \leq q \leq \min[\varepsilon + G(S), |Y| - k - 1]$, where $\varepsilon = c - a$, and (2) for C-terminal seed: $\max[k, |Y| + (\varepsilon - G(S)) - 1] \leq q \leq \min[|Y| + k - 1, |Y| + (\varepsilon + G(S)) - 1]$, where $\varepsilon = d - b$. Note that the query embedding position $q$ is computed relative to the original query sequence positions. For example, if the N-terminal seed is inserted at the beginning of the query, $q$ is then $-k$ in order to preserve the original query sequence positions in the alignment. Recall that the region of interest starts immediately after the seed and in this way we can preserve the original positions for the subsequent computations. The insertion positions for the C-terminal seeds are also represented similarly. For C-terminal calculations, if $\max[k, |Y| + (\varepsilon - G(S)) - 1]$ is larger than $\min[|Y| + k - 1, |Y| + (\varepsilon + G(S)) - 1]$, no C-terminal seed is inserted. The idea of maximum gap has been described previously for connecting HSPs with gaps(9).

**Step 3. Generate final alignments with a seed.** For each query position $q$ identified in step 2, we perform an alignment with the seed $S$ inserted in the respective position to generate the final alignments. The final alignments are generated by running dynamic programming starting at the end position of the seed, ($|S|-1$, q), and proceeding to ($|X|-1$, $|Y|-1$). Since we are working with highly divergent sequences producing low-identity alignments, it is reasonable to consider the scenario that a longer alignment with lower score can be biologically more meaningful than a shorter alignment with higher score(11). Motivated by this observation, during the alignment, we adjust an alignment score with respect to the length of the alignment as follows:

$$s_a(a,b) = \begin{cases} s_c(a,b) \log a & if a > 1 \\ s_c(a,b) & otherwise \end{cases}$$

where $(a,b)$ is a cell in the dynamic programming matrix, and $s_c(a,b)$ and $s_a(a,b)$ is the score before and after adjustment, respectively. Note that $a$ represents the alignment length at position $(a,b)$ in the dynamic programming matrix. If we have the best score at $(a,b)$, we have the final alignment ($x_{|S|-1,a}$, $y_{q,b}$).

**Step 4. Filter out non-significant alignments.** Not all alignments produced from the previous step are informative. In this step, we prune out insignificant alignments using the metrics, % *coverage* and *pairwise identity*. Given an alignment ($x_{a,b}$, $y_{c,d}$), the % coverage of the alignment to a target sequence is calculated as follows:

$$Coverage = \begin{cases} \dfrac{b+1}{|X|} & if N-terminal\ seed \\ \dfrac{|X|-a}{|X|} & otherwise \end{cases} \%$$

where $X$ is a target sequence. The pairwise identity considered here is the identity of the alignment excluding the matches in a seed: i.e. *pairwise identity*= $\dfrac{\#\ of\ matches}{\alpha} \times 100$, where $\alpha$ = # of matches + # of mismatches + # of gaps in an alignment excluding the seed. If the pairwise identity and % coverage of a final alignment are greater than the thresholds, *minimum identity* and *minimum coverage*, the alignment is returned to the user.

### *Preparation of function or structure-specific PSSM set*

To generate the PSSM set for a specific protein function or structure fold, we first collected protein sequences which are known to be related to the function or structure of our interest. For the PSSM set, we generated PSSMs with the collected sequences or the sequences expanded from the collected sequences using PSI-BLAST(24). For expansion, each collected sequence is searched against NCBI NR database by PSI-BLAST (with the option of -e $1 \times 10^{-3}$ –h $1 \times 10^{-6}$). Among the returned sequences, we filtered out the sequences whose pairwise identity to the query sequence is more than 90% and redundant sequences in the set. And, for PSSM generation for those expanded sequences, PSI-BLAST (with the option of –h $1 \times 10^{-6}$) was used again. All PSSM sets which have been used for our test will be provided upon request.

### *Experimental setup and data sets*

Both GDDA-BLAST and Ada-BLAST were implemented in C, and compiled for both UNIX and Windows environments. GDDA-BLAST utilizes rps-BLAST in NCBI BLAST 2.2.15 package to compute the alignments. In order to validate the approach, we tested both for the execution time and the accuracy. The execution time experiment was conducted in a dedicated machine with 1.8GHz Intel Core$^{TM}$ 2 duo processor and 2GB memory running Windows Vista. The experiment for accuracy was performed on a server with eight Dual-core 2.4 GHz AMD Opteron processors and total of 32G memory running Linux. Note that for the execution time experiment we used a less-equipped dedicated machine instead of the server shared by others in order to measure the execution time more accurately.

## Results
### *Execution time and accuracy*

To compare the execution time of GDDA-BLAST and Ada-BLAST, we ran both methods with 602 query sequences randomly chosen from the SABmark twilight zone set (27) and 51 target sequences randomly selected from the CDD database. Figure 7 shows the per-query execution time when a given query is run against the 51 PSSMs in the library. The lengths of the 602 query sequences range from 34 to 759 amino acids. Note that the running time of GDDA-BLAST increases linearly as a function of query sequence length. Conversely, Ada-BLAST shows much better scalability with respect to the size of query because it inserts a seed only at the positions where the seed is likely to be extended. Moreover, the performance gain is maximized when the two sequences compared are of low identity because the number of likely seed-insertion positions is limited. This makes Ada-BLAST an attractive alternative for the alignment of highly divergent sequences. Overall, Ada-BLAST is 19.3 times ($\pm$ 15.29 S.D.) faster than GDDA-BLAST on average while it achieves more than 100 times speed-up in many occasions.

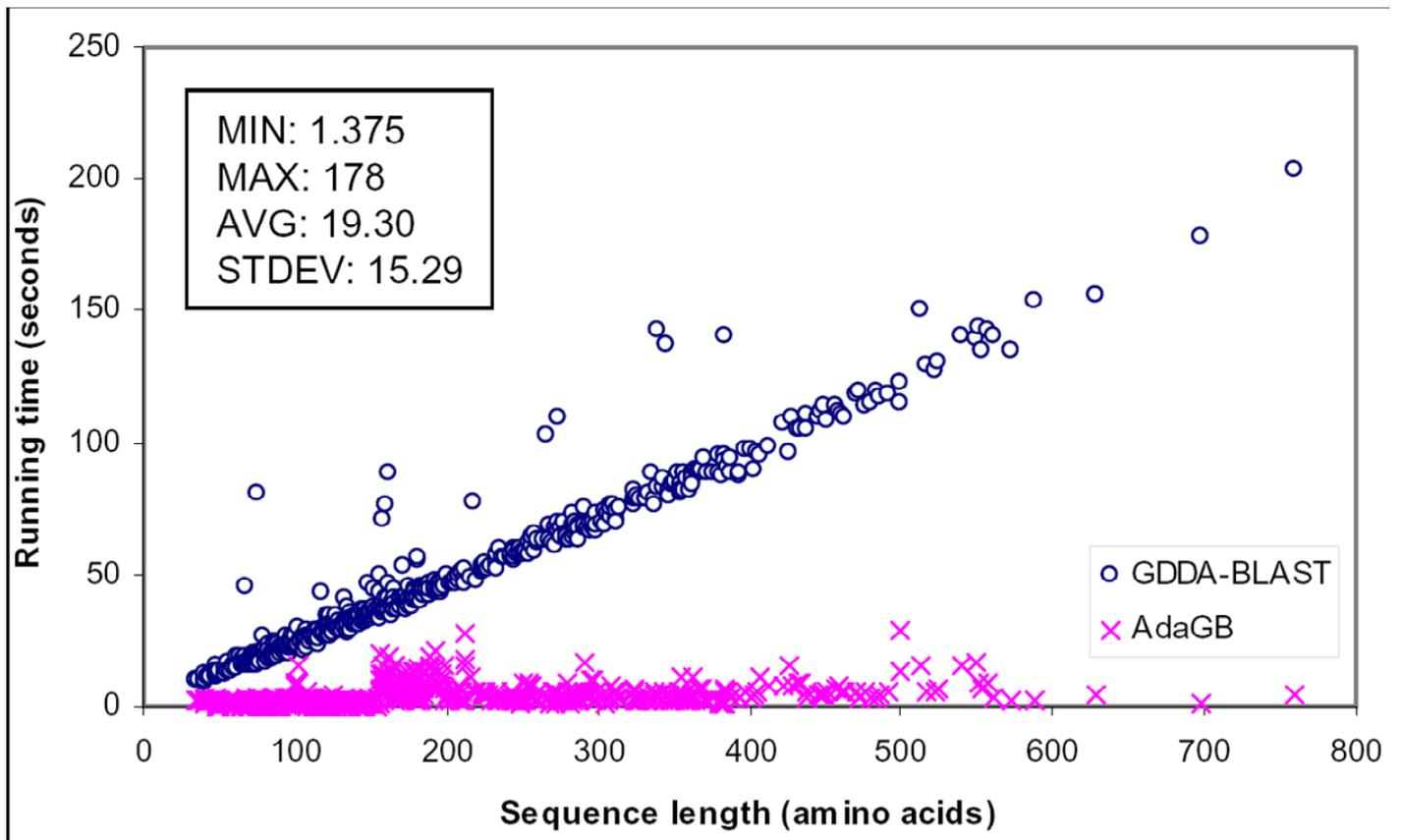

**Figure 7 Running time of GDDA-BLAST and Ada-BLAST.** Comparison of running time for 620 query sequences against 51 target sequences. The numbers in a box represent how much Ada-BLAST is faster than GDDA-BLAST.

We performed Receiver Operating Characteristic (ROC) curve analysis [4] to see how accurately GDDA-BLAST and Ada-BLAST recognize queries in the same group as related sequences (data not shown). ROC curve shows the sensitivity of each method at different false positive rates. Sensitivity and false positive rate are calculated as follows: $Sensitivity = \frac{TP}{TP+TN}$, $Specificity = \frac{TN}{TN+FP}$, *False Positive Rate = 1 - Specificity*, where *TP* is # of true positives, *TN* is # of true negatives and *FP* is # of false positives.

In this evaluation, we used 534 sequences out of 61 sequence groups randomly selected from the SABmark Twilight zone set, as described previously(1). Pearson's correlation is used for GDDA-BLAST and Ada-BLAST and sequences with *K* highest Pearson's correlations to the query are returned. The difference in accuracy of GDDA-BLAST when compared with Ada-BLAST is negligible.

*Characterization of Secondary Structural Elements*

In order to determine the information content contained in a pure population of embedded alignments we analyzed the structurally resolved (X-ray Crystallography) transmembrane protein, Bovine Rhodopsin (PDB: 1F88)(28). Figure 8A depicts the output of rps-BLAST (e-value threshold 0.01) for the domain architecture of 1F88. Notably, rps-BLAST returns overlapping alignments for 5 different PSSMs defined as Serpentine type 7TM domains. Our theories on structurally/functionally related PSSM libraries predict that additional information below accepted statistical thresholds can be utilized to define, with higher resolution, domain boundaries and secondary structural elements.

**A** Query PDB: 1F88A Chain A, Crystal Structure of Bovine Rhodopsin

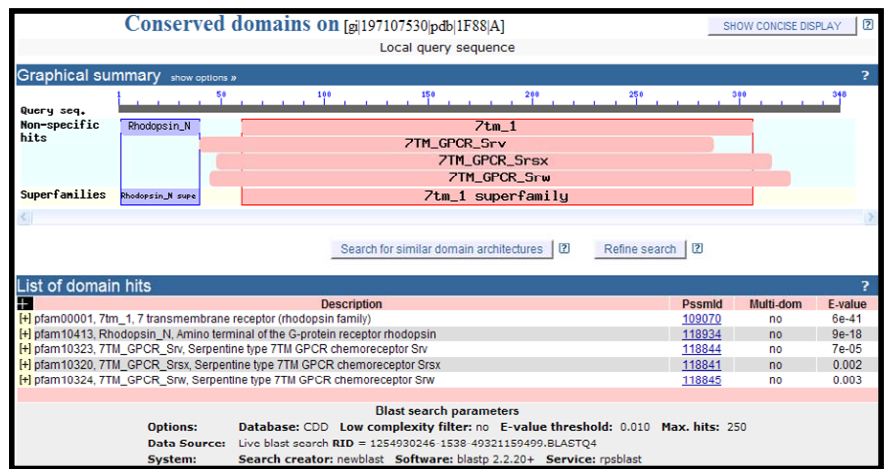

**B**

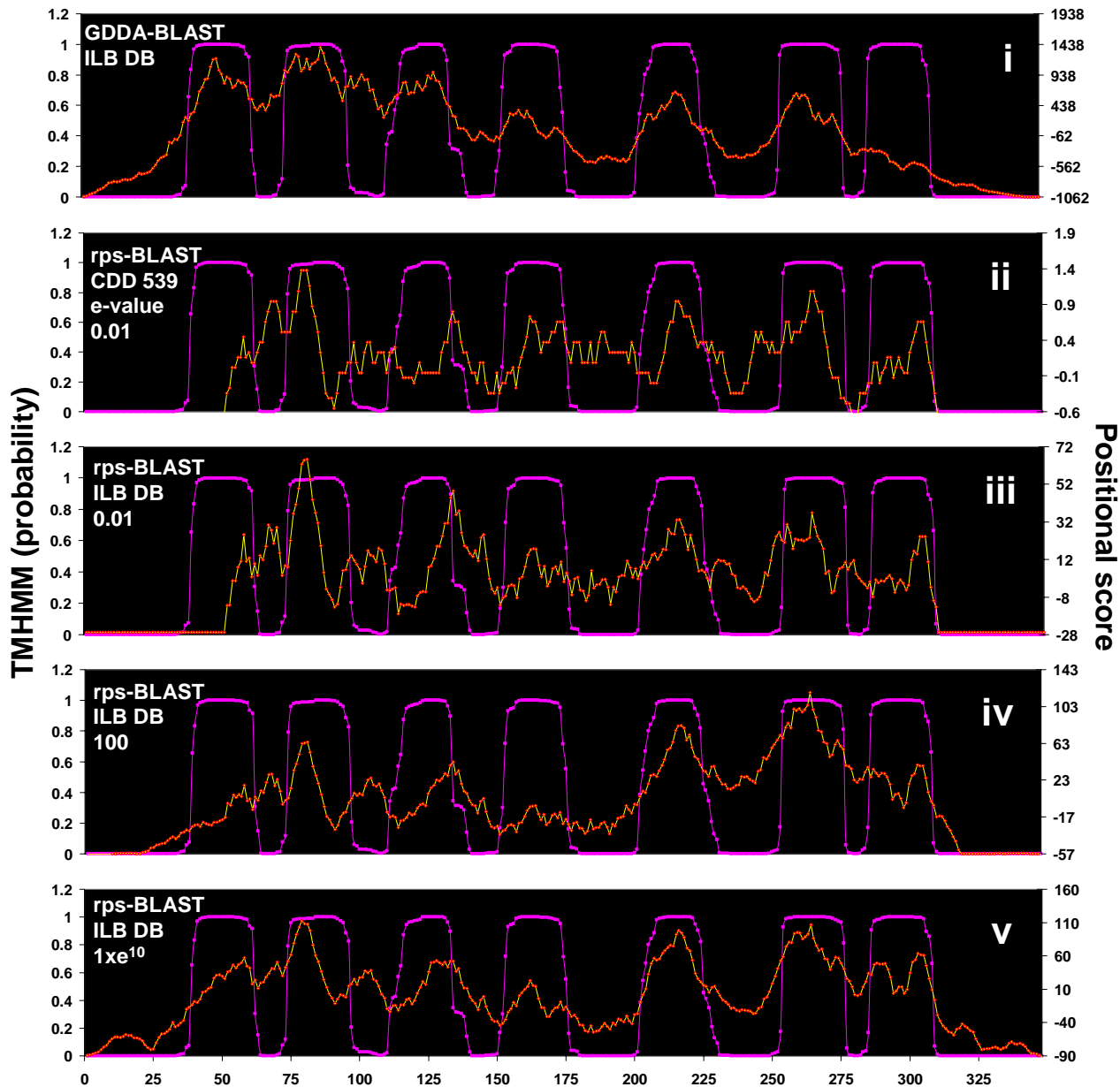

**Figure 8 Comparisons of Positional Scores and Markov Models.** (a) Depicts the output of rps-BLAST (e-value threshold 0.01) for the domain architecture of 1F88. (b, i-v) Graphs show alignment results which were smoothed by adjacent averaging with a 7 amino acid window. For each case the transmembrane probability determined by TMHMM is shown on the left axis (magenta). The right axis represents a positional score for Ada-BLAST and rps-BLAST conditions.

This hypothesis was tested and the performance was evaluated against rps-BLAST and Hidden Markov Models (TMHMM)(29). For this experiment we considered multiple running conditions for rps-BLAST: (i) run rps-BLAST against the CDD database, and among the alignments reported, consider only those PSSMs which we have previously annotated as transmembrane by keyword (CDD 539), (ii) run rps-BLAST against a database of PSSMs (n= 24,378) derived from expanding all of the sequences contained in the original 539 PSSMs using PSI-BLAST (integral lipid-binding database, ILB DB) (see Methods), and (iii) run rps-BLAST against these databases and slide the e-value threshold to less statistically significant levels (0.01, 100, or $1xe^{10}$).

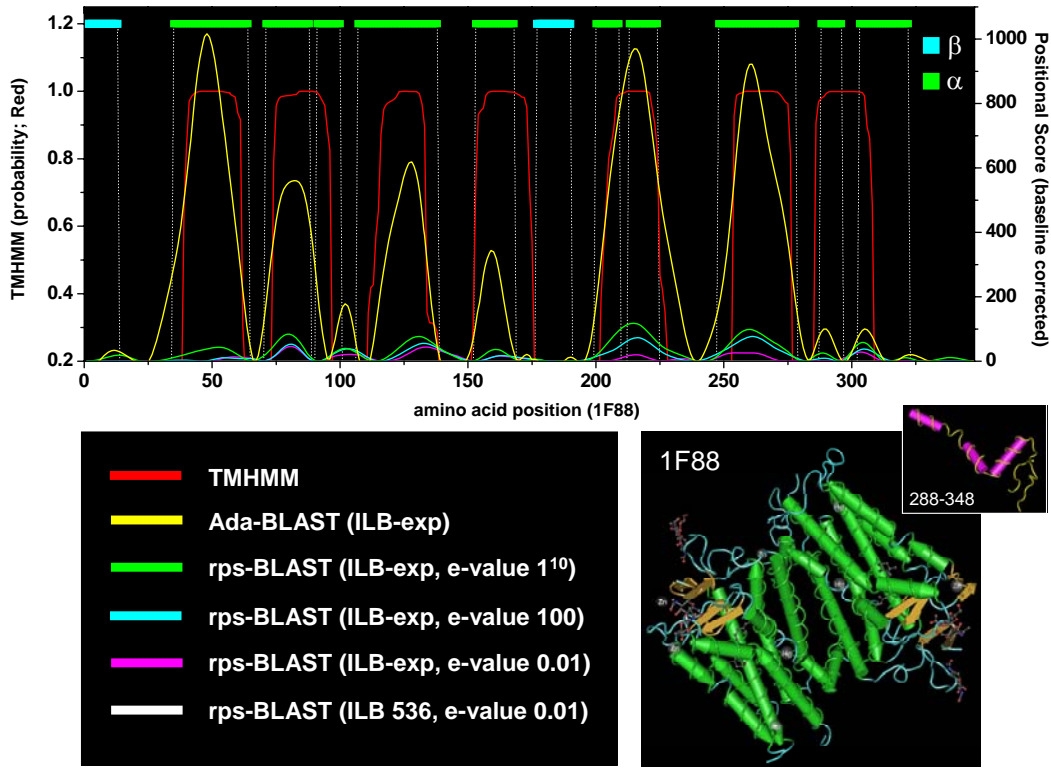

The graphs in Figure 8b show the alignment results which were smoothed by adjacent averaging with a 7 amino acid window. For each case the transmembrane probability determined by TMHMM is shown on the left axis (magenta). The right axis represents a positional score for Ada-BLAST and rps-BLAST conditions. The positional score was quantified in the following manner. For each positive PSSM, the alignment boundaries as determined by the overlapping alignments obtained from either Ada-BLAST or rps-BLAST. These regions were extracted and realigned by the Smith-Waterman algorithm with a BLOSUM62 and BLOSUM45 substitution matrix. Identical residues were scored value=2 and those which were positive (non-identical but conserved) were scored value=1. This process was repeated for all positive PSSMs and the results summed for each amino acid in the protein. Currently, we do not know what substitution matrix is the most optimal so we averaged the positional scores derived from BLOSUM62 and BLOSUM45. The positional results in Figure 8B (i-v) were normalized to zero by subtracting the average positional score across the protein length from each point and then each amino acid position was subjected to adjacent averaging. These results demonstrate that: (i) using the expanded ILB DB increases the signal-to-noise ratio using either rps-BLAST or Ada-BLAST, (ii) Ada-BLAST has a ~10-fold increase in signal compared to the largest results obtained from rps-BLAST ($1xe^{10}$), and (iii) the positional data, even at the highest statistical limit of rps-BLAST tested with the expanded ILB DB, accord with results obtained from TMHMM.

**Figure 9 Characterization of Membrane Spanning Regions.** This graph shows the performance of Hidden Markov Models (TMHMM), rps-BLAST, and Ada-BLAST in the determination of the membrane spanning domains in Bovine Rhodopsin (Teal= Beta pleated sheets, Green=helices, loops not shown). This protein was analyzed with an expanded set of PSSMs representing a large variety of transmembrane domains (~20K PSSMs). Compared with rps-BLAST, Ada-BLAST is more refined with respect to the annotation of alpha-helices. Moreover, this data demonstrates that less statistically valid alignments (e.g. e-value 0.01 vs 100) are still informative for detecting domain boundaries and outperform lower thresholds. The full-length structure of Rhodopsin is shown (dimer) as well as an inset of the C-terminus which is composed of three small helices with the last one folding parallel with the membrane (it itself is not believed to be transmembrane).

It is reasonable to consider that amino-acids within transmembrane spanning helices will be more conserved than the intervening loop residues. Support for this hypothesis is presented in Figure 9, wherein we report our results when compared to the known structural elements of 1F88 as obtained from X-ray crystallography(28). The full-length structure of Rhodopsin is shown in the bottom right (dimer) as well as an inset of the C-terminus. For this analysis, we used Origin Lab 7.5© to perform smoothing (Fourier-transform point=8) and discontinuous baselining of the positional data. This correction was performed by baselining to every local minimum across the entire curve. The structural features are annotated with droplines (Cyan= Beta pleated sheets, Green=helices, loops not shown). In all cases, the curves obtained from the positional data can be correlated to the known structural elements.

While none of the prediction methods accurately model all of the crystal structure, we observe several interesting features. For example, several of the membrane-spanning helices are interrupted by loop regions that are not identified by TMHMM. Indeed, the C-terminus of 1F88 contains 3 small helices, the last of which is a bent-helix believed to be parallel to the membrane (aa 288-348, Figure 9 *inset*). Both rps-BLAST and Ada-BLAST detect these smaller helices with Ada-BLAST having the highest signal. Another region of interest is contained between aa 91-111, which is a loop in the crystal structure, but is predicted to be a short helix by rps-BLAST and Ada-BLAST. We theorize that this loop may be, under native conditions, a bent-helix similar to other regions in the protein.

### *Classification based on Hierarchical Clustering*

We hypothesize that the alignment information described above can be used to classify protein structure/function when encoded into a phylogenetic profile. In general, similarity measurements between two protein sequences are done by directly aligning the two sequences one against the other. However, GDDA-BLAST and Ada-BLAST do not compare the two query sequences directly. Instead, we compute phylogenetic profiles for each query sequence which provides a quantitative platform for sequence comparison(1;2;4). In addition, we have made several modifications to rps-BLAST, such that its data can be encoded into phylogenetic profiles.

A phylogenetic profile of a query is a vector of length equal to the number of target sequences such as $A = (b_1, b_2, \ldots, b_m)$ where $m$ is the number of target sequences. Each dimension of the vector, $b_j$, represents the alignment score for the query sequence against the $j^{th}$ target sequence and computed as $b_j = h_j \times s_j \times v_j$ where $h_j$ is the number of alignments between the query sequence and the target sequence after filtering (in step 4), $s_j$ is the average %identity of the alignments, and $v_j$ is the maximum coverage among the alignments. Note that $h_j$ can be up to the length of the query sequence (i.e., #of embedded seeds) because we align the embedded sequence of the query against the target. In the case of rps-BLAST, the calculations are the same except for the presence of embedded sequence.

We use Pearson's correlation coefficient to measure the similarity between two query profiles. Pearson's correlation coefficient between two profiles $A$ and $B$ is computed as follows(30):

$$S(A, B) = \frac{1}{m} \sum_{i=1}^{m} \left( \frac{A_i - A_{offset}}{\Phi_A} \right)\left( \frac{B_i - B_{offset}}{\Phi_B} \right)$$

where $m$ is the length of the vector, $A_{offset}$ is the average of $A$, and $\Phi_A = \sqrt{\sum_{i=1}^{m} \frac{(A_i - A_{offset})^2}{m}}$.

**Figure 10 Classification based on Heirarchical Clustering.** 74 sequences representing multiple classes of transmembrane containing proteins were hierarchically clustered by the microarray analysis algorithm Cluster (settings: Complete Linkage and Correlation (centered) similarity metric) and visualized utilizing the Treeview algorithm. Red lines represent the correlation scores derived from the analysis. Alignments for the 24,378 ILB DB PSSMs were derived by either (a) Ada-BLAST, or (b) rps-BLAST.

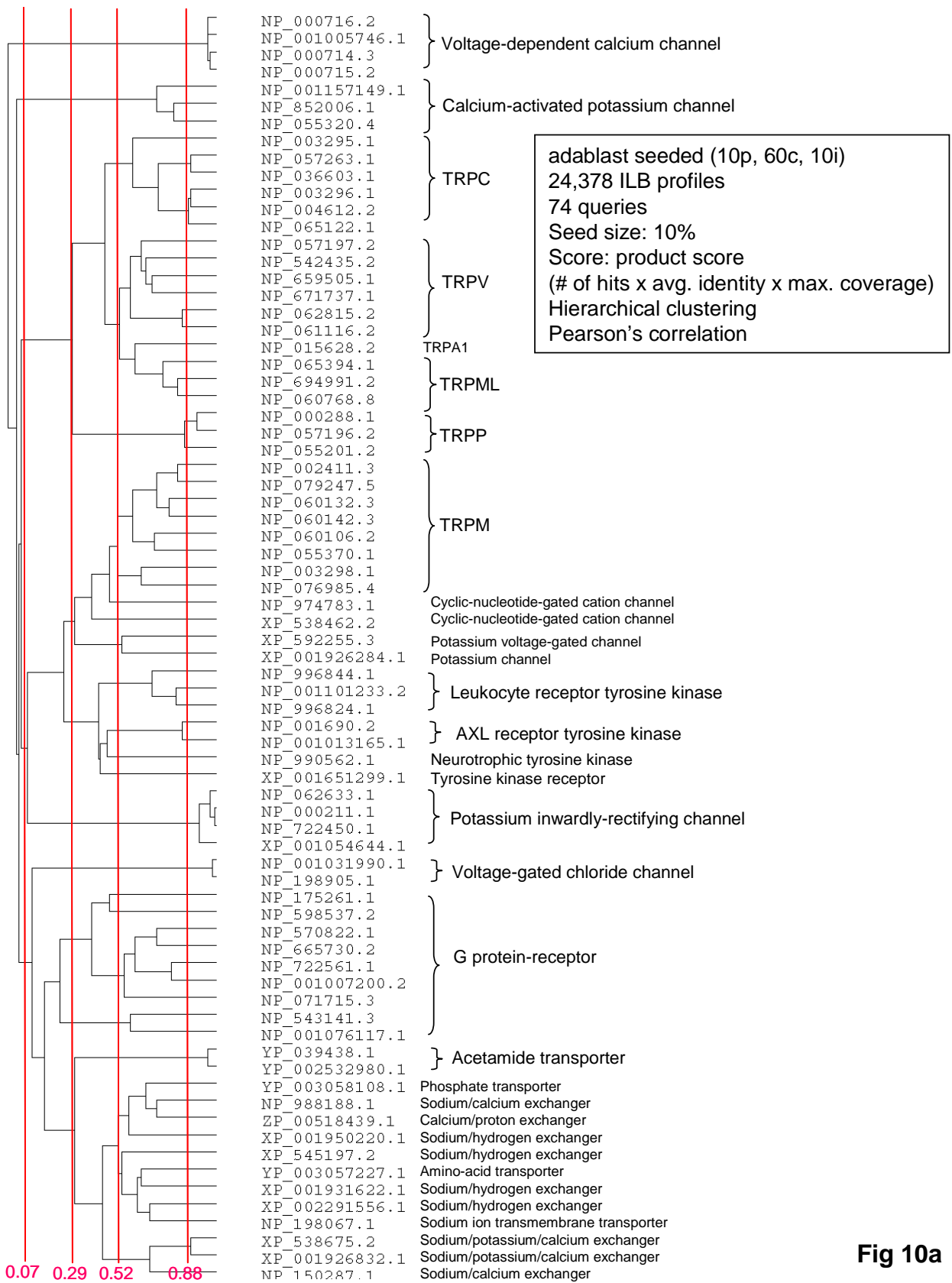

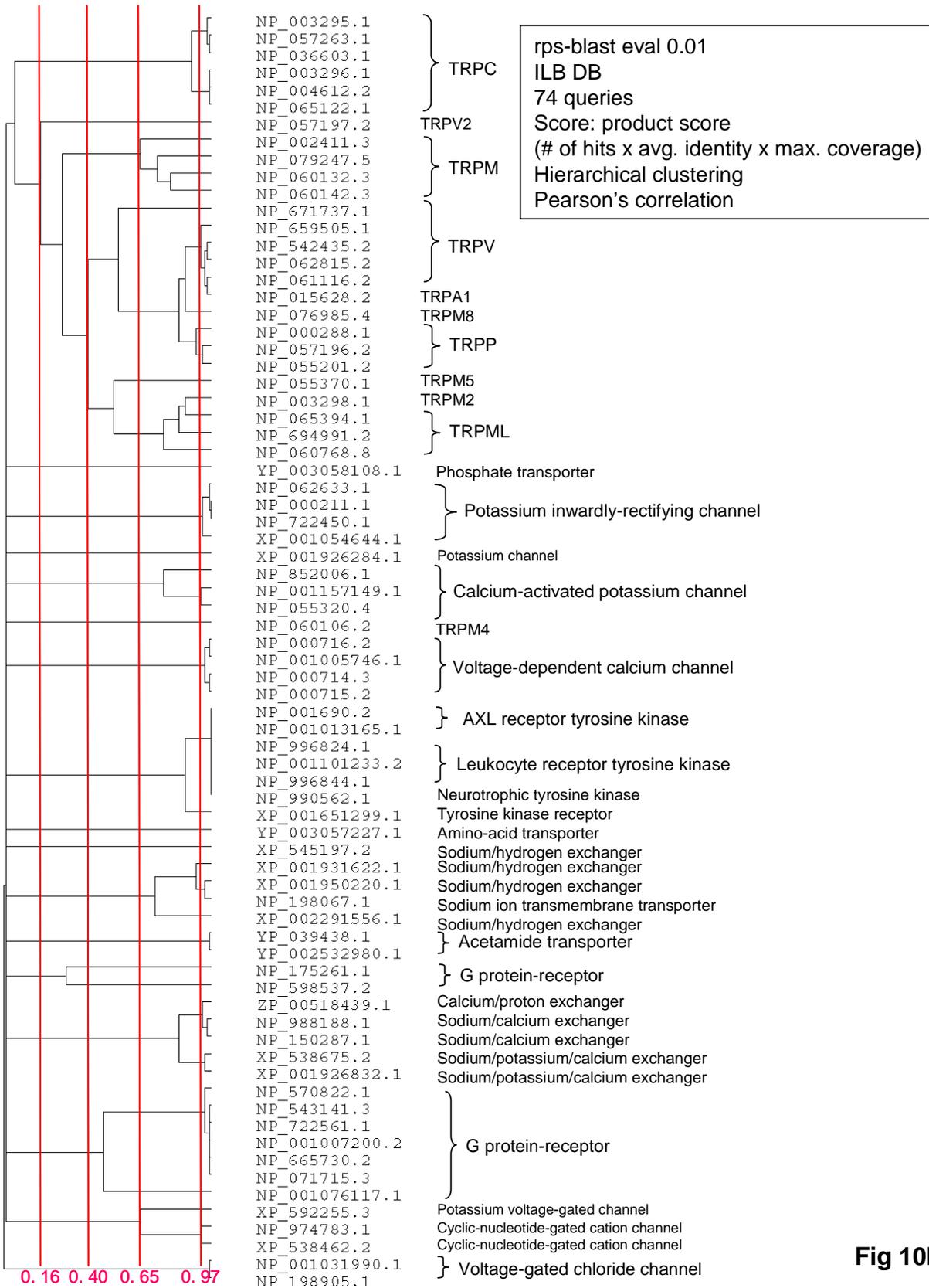

**Fig 10b**

To test the efficacy of phylogenetic profiles built using either a pure population of embedded alignments or alignments over a range of rps-BLAST e-value thresholds, we curated a set of transmembrane containing proteins from a range of different protein families (voltage-gated $Ca^{2+}$, $K^+$, and $Cl^-$ channels, calcium-activated $K^+$ channels, cyclic-nucleotide gated channels, transient receptor potential channels (TRPs), receptor tyrosine kinases, G-protein coupled receptors (GPCRs), transporters, and exchangers). Each of these 74 sequences was analyzed using the expanded ILB DB with Ada-BLAST and rps-BLAST. The results from this analysis were encoded into a NXM matrix using the scoring scheme ([# of hits] x [mean % identity] x [max % coverage]) and then subjected to hierarchical clustering. In Figure 10 a-b, we report the conditions that achieve the highest degree of classification (Ada-BLAST and rps-BLAST e= 0.01 respectively).

While neither of these classifications are perfect, both are robust for pairing related sequences; rps-BLAST has the highest Pearson's correlation, while Ada-BLAST has a more robust topology. Specifically, Ada-BLAST classifies the TRP channels, GPCRs, and transporters/exchangers into their correct group, while rps-BLAST does not. In data not shown, we observe that both the Pearson's correlation values and the overall topology is compromised when measured at e= 100 and e= $1xe^{10}$ in rps-BLAST. Taken together, these data demonstrate that a pure population of embedded alignments can differentiate between proteins having similar structure and diverse function given the appropriate PSSM database.

**Discussion**

In this study, we provide evidence for our theories that: (i) sequence embedding amplifies low-identity alignments and that these alignments are distinct from those derived by simply scaling e-value thresholds, (ii) low-identity alignments contain information which can inform protein structure/function annotation before and after encoding into a phylogenetic profile, and (iii) PSSM libraries which are constructed from key-word searches of the CDD database, expanded using PSI-BLAST and then made into a specific database can be used to classify proteins based on their structure/function. There are several implications which can be drawn from these findings.

The biochemical characterization of membrane spanning proteins is challenging. Likewise, structural studies of this protein class can be fraught with artifacts introduced by crystallization and/or lack of appropriate co-factors (e.g. lipids). Our results demonstrate that signals derived by Ada-BLAST can provide structural information that is distinct from Markov Models of transmembrane regions. It is tantalizing to consider that these measurements may isolate potential discrepancies in protein structure due to non-native conditions. Importantly, when encoded into a phylogenetic profile, these same data can be utilized to functionally classify transmembrane containing proteins.

The power underlying protein phylogenetic profiles is centered on the construction of PSSMs representing specific folds and/or activities. Indeed, results from this manuscript and also our companion study (Ko et al, Physics Archives November 2009) demonstrate that PSSMs libraries generated using specific folds can accurately identify homologous folds. Moreover, PSSM libraries generated for a specific activity can accurately identify homologous functions in proteins of diverse structure, as well as differentiating activity within a specific fold. Within the present study we generated transmembrane PSSMs curated by key-word searches; however, it is likely that additional refinement of this set (e.g. to remove non transmembrane regions from PSSM) will lead to an increase in the signal/noise ratio, better annotation of transmembrane spanning domains, secondary structure prediction, and more robust classification. We are actively pursuing our working hypothesis that this approach will work for any class of protein domains.

Our results support the idea that statistical thresholds are often too stringent in domain detection algorithms. For example, rps-BLAST does not report a channel domain alignment in human TRPV channel (gi|22547180) at statistical limits. In the present study we found that additional information contained in alignments well below accepted statistical thresholds can be utilized to inform domain boundaries and secondary structural elements. However, this was not the case when these same data were hierarchically clustered. Future analysis on sufficiently large datasets is required to identify and optimize the multiple variables which can identify highly-divergent yet informative alignments. Nevertheless, we propose that there is a wealth of information below statistical values which can aid researchers in annotating protein structure/function.

In conclusion, we propose that future work aimed at (i) creating comprehensive and refined PSSM libraries and (ii) exploration of sequence embedding at the level of the PSSM (COBBLER, (25)) and within the query (Ada-BLAST), will exponentially increase the functional annotation of all classes of proteins across taxa. Such an advance would have broad impacts on human health and disease, as well as basic science endeavors. Indeed, since Ada-BLAST performs in the "twilight zone" of sequence similarity, this approach can be harnessed to decode the most challenging protein datasets, and scaled up to screen proteomes and the vast quantities of sequences being obtained from metagenomic studies. Outside of biological questions, the theories behind these algorithms are likely to have applications in many other fields that use pattern-based prediction algorithms.

**Acknowledgements-** This work was supported by the Searle Young Investigators Award and start-up money from PSU (RLP), NCSA grant TG-MCB070027N (RLP, DVR), The National Science Foundation 428-15 691M (RLP, DVR), and The National Institutes of Health R01 GM087410-01 (RLP, DVR). This project was also funded by a Fellowship from the Eberly College of Sciences and the Huck Institutes of the Life Sciences (DVR) and a grant with the Pennsylvania Department of Health using Tobacco Settlement Funds (DVR). The Department of Health specifically disclaims responsibility for any analyses, interpretations or conclusions. This work was also supported in part by Korea University Grant, the Second Brain Korea 21 Project Grant, Korea Research Foundation Grant funded by the Korean Government (MOEHRD) (KRF-2008-331-D00481), and Korea Sciences and Engineering Foundation (KOSEF) grant funded by the Korea government (MEST)(R01-2008-000-20564-0). We would especially like to thank Jason Holmes and the CAC center for their superior support. We would also like to thank Drs. Robert E. Rothe, Jim White, Kenji Cohan, Max Hong, Barbara Van Rossum, and J. Hendrix for creative dialogue.